\documentclass[12pt]{iopart}
\usepackage{graphicx}

  \expandafter\let\csname equation*\endcsname\relax
  \expandafter\let\csname endequation*\endcsname\relax 

 \usepackage{amssymb}
 \usepackage{amsbsy}
 \usepackage{amsfonts}
\usepackage{amsmath}

\begin{document}

\title{Magnetized strongly coupled plasmas and how to realize them in a dusty plasma setup}

\author{M Bonitz$^1$, H K\"ahlert$^2$, T Ott$^1$, and H L\"owen$^2$}

\address{$^1$ Christian-Albrechts-Universit\"at zu Kiel, Institut f\"ur Theoretische Physik und Astrophysik, Leibnizstr. 15, 24098 Kiel, Germany}
\address{$^2$ Heinrich-Heine Universit\"at D\"usseldorf, Institut f\"ur Theoretische Physik II: Weiche Materie, Universit\"atsstr. 1, 40225 D\"usseldorf, Germany}
\ead{bonitz@physik.uni-kiel.de}
\begin{abstract}
Strongly coupled plasmas in which the interaction energy exceeds the kinetic energy play an important role in many astrophysical and laboratory systems including compact stars, laser plasmas and dusty plasmas. They exhibit many unusual collective properties, such as liquid or crystalline behaviour, peculiar oscillation spectra and transport properties. Recently, strongly coupled plasmas were studied in the presence of a strong magnetic field by computer simulations, and strong modifications of their transport properties and oscillation spectra were observed. While strong magnetization is common in stellar systems it is practically impossible to achieve in complex plasmas due to the large mass of the dust particles. Here we discuss a recently demonstrated approach to achieve very strong ``magnetization'' by a rotation of the neutral gas, and we present new results for macroscopic two-dimensional systems.
\end{abstract}

\submitto{\PSST}

\maketitle

\section{Introduction}\label{s:intro}
Strongly coupled plasmas in which the Coulomb interaction energy per particle exceeds the thermal energy are of high interest in dusty plasmas, trapped ions, colloidal systems, quantum confined semiconductor structures as well as in diverse astrophysical environments. In recent years much attention has been devoted to study the structural properties, oscillation spectra and transport properties of these systems both theoretically and experimentally. A particularly detailed analysis was performed for colloidal and dusty plasmas, e.g. \cite{ivlev2012,bonitz2010}, where one takes advantage of the large particle size (micrometers) that allows for single-particle resolution and optical diagnostics. Therefore, the latter plasmas have been successfully used as a unique ``laboratory for strong correlations'' in Coulomb systems in general \cite{bonitz2010} allowing, for example, to make predictions of important structural and dynamical properties of such exotic objects as white dwarf stars or neutron stars, which are not directly accessible to experimental studies.

Many strongly correlated plasmas are, in addition, subject to a strong magnetic field which may significantly alter the plasma behaviour. Examples are the electron-hole plasma in semiconductor quantum wells (quantum Hall systems), magnetic target fusion scenarios or the plasma in the outer layers of neutron stars \cite{potekhin2010}. Recent theoretical studies showed that the magnetic field drastically alters the particle and energy transport in these systems, in particular the diffusion coefficient \cite{ott2011}, see Sec.~\ref{s:magnetization} and the collective oscillation spectrum \cite{bonitz_prl10,ott_prl12}. It is tempting to directly verify these predictions in the laboratory with dusty plasmas, making use of the above mentioned unique spatial and temporal resolution. Unfortunately, due to the large particle size, it is practically impossible to magnetize a dusty plasma, see Sec.~\ref{s:magnetization}. We have recently presented a solution to this dilemma: the key idea is to put the dust particles into rotation via a rotation of the neutral gas. Then, in the rotating frame, the dust particles are subject to the Coriolis force that is equivalent to the Lorentz force of a magnetic field. A brief derivation and proof of principle experiment was reported in~\cite{kaehlert2012} for a system of 4 dust particles in a plane.
In this paper we present a more detailed derivation of the equations of motion of a rotating dusty plasma, extend the analysis to macroscopic 2D systems and analyze the achievable ``magnetization'', see Sec.~\ref{s:rotation}. Before that we introduce, in Sec.~\ref{s:magnetization}, the relevant parameters of a strongly correlated magnetized plasma, discuss important special cases and the main effects of the magnetic field.

\section{Magnetization effects in strongly coupled plasmas}\label{s:magnetization}
\begin{figure}
\includegraphics[width=3.5cm]{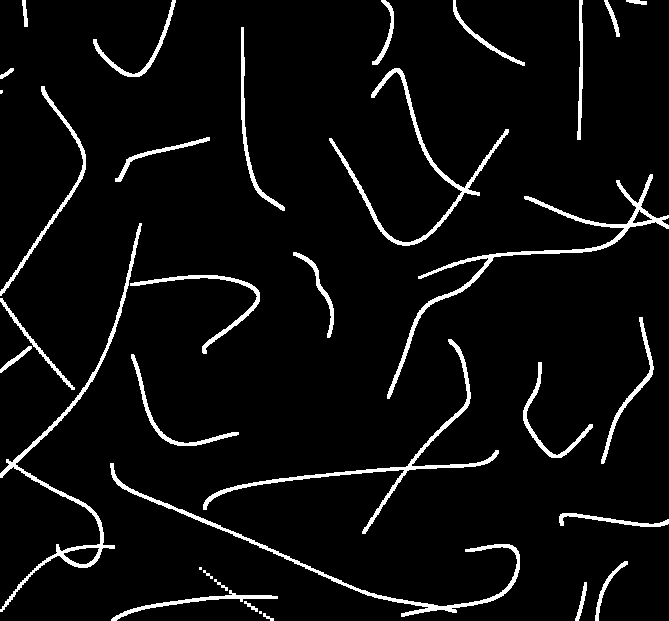}
\includegraphics[width=3.5cm]{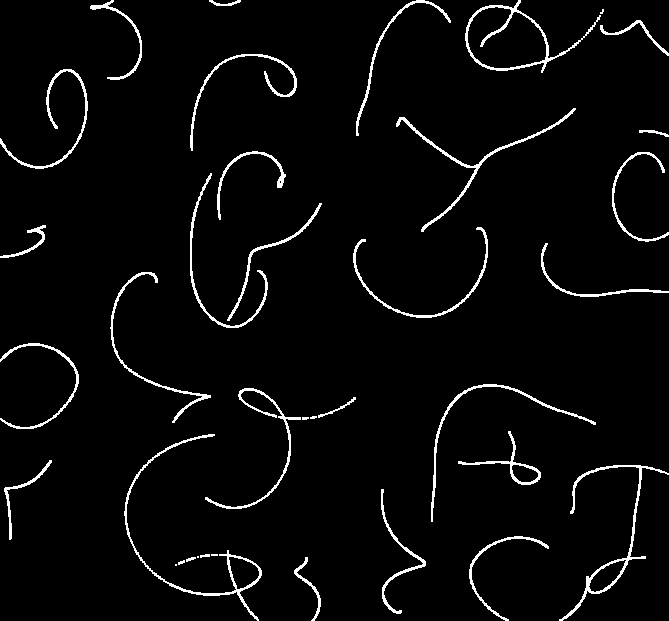}
\includegraphics[width=3.5cm]{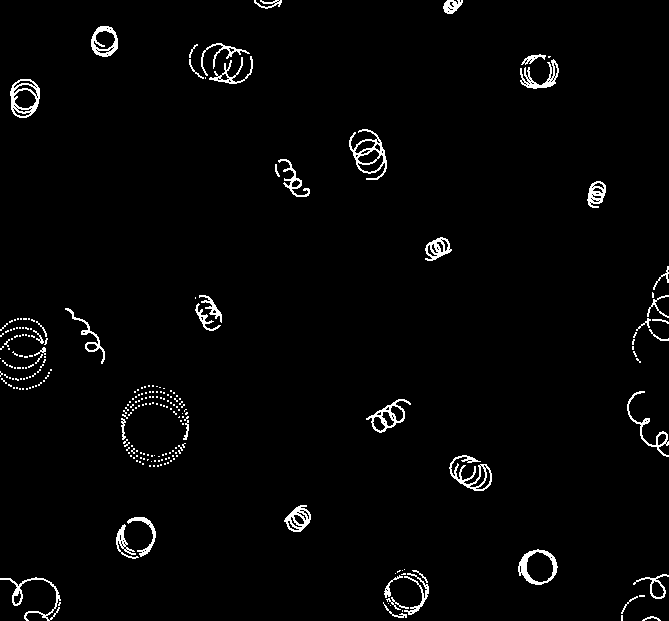}\\[0.5ex]
\includegraphics[width=3.5cm]{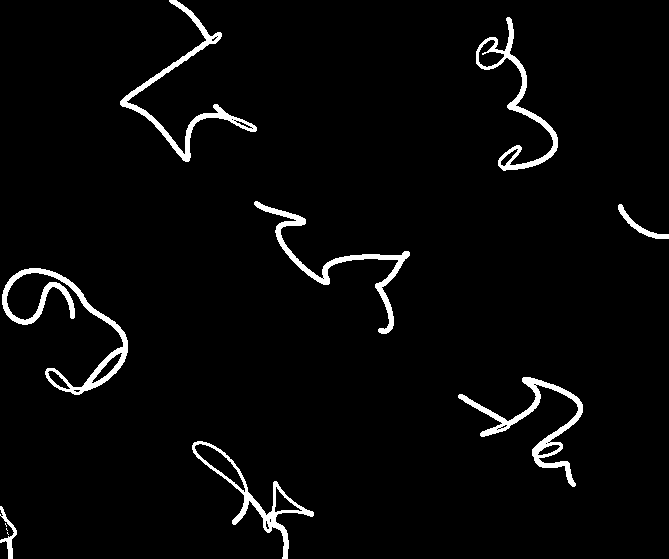}
\includegraphics[width=3.5cm]{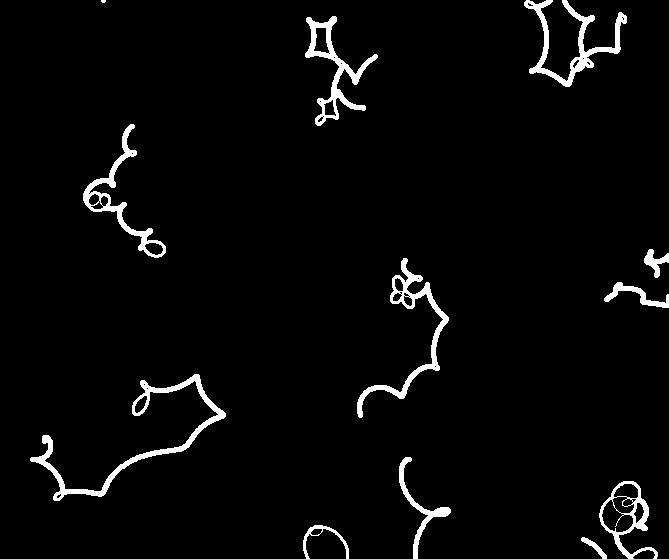}
\includegraphics[width=3.5cm]{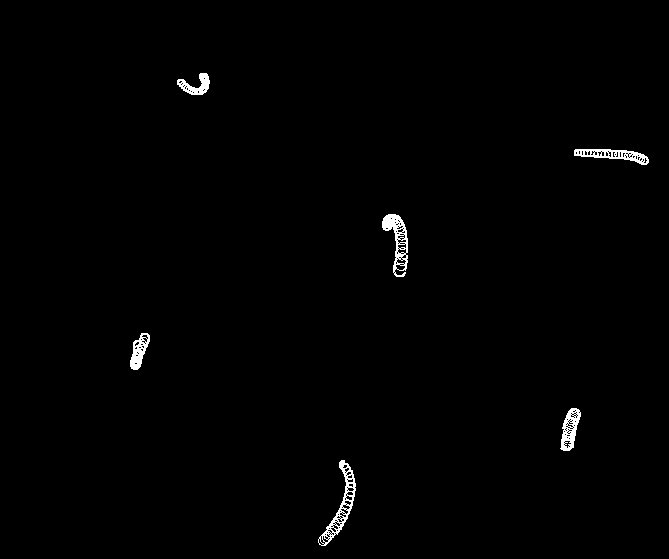}
\caption{Trajectories of a moderately ($\Gamma=2$, top, five plasma periods) and strongly coupled ($\Gamma=100$, bottom, fifty plasma periods) one-component 2D plasma for different magnetic field strengths. Left column: $\beta = 0$, middle column: $\beta = 1$, right column: $\beta = 4$. Results of MD simulations.}\label{fig:traject}
\end{figure}

A classical strongly coupled Coulomb system in thermodynamic equilibrium is described by a single dimensionless parameter, the coupling parameter
\begin{equation}\label{eqn:gamma}
 \Gamma=\frac{Q^2/4\pi\epsilon_0}{a_\text{ws}k_BT }.
\end{equation}
This assumes that the dynamics can be described by that of an effective one-component system (OCP) where the second component assures neutralization (and possibly screening of the interaction) but does not dynamically participate. This is well fulfilled in the case of large mass asymmetry of the components, such as in dusty plasmas (if ion streaming is not important) or in dwarf stars and the envelope of neutron stars and will be at the basis of the present analysis. Thus, eq.~(\ref{eqn:gamma}) contains only properties of the heavy component: the particle charge $Q$, the mean inter-particle distance (Wigner-Seitz radius) $a_\text{ws}$ and the temperature $T$ that is related to the thermal velocity $v_T=(k_BT/m)^{1/2}$, where $m$ is the mass. A second quantity that characterizes the strength of the Coulomb interaction is the plasma frequency (dim$=2, 3$ is the dimensionality),
\begin{equation}\label{eqn:plasmafreq}
\omega_\text{p}={\rm dim}^{1/2}\sqrt{\frac{Q^2/4\pi\epsilon_0}{ ma_\text{ws}^3}}.
\end{equation}
Typical correlation effects are depicted in figure~\ref{fig:traject}, where we show the trajectories of several particles computed from a molecular dynamics (MD) simulation. For moderate coupling ($\Gamma=2$, top left) free flight periods are occasionally interrupted by collisions that bend the trajectories. In contrast, for strong coupling ($\Gamma=100$, bottom left), the interaction is so strong that particles are trapped in local  minima of the total potential energy (``caging'') exhibiting localized oscillations.

\begin{figure}
\rotatebox{-90}{\includegraphics{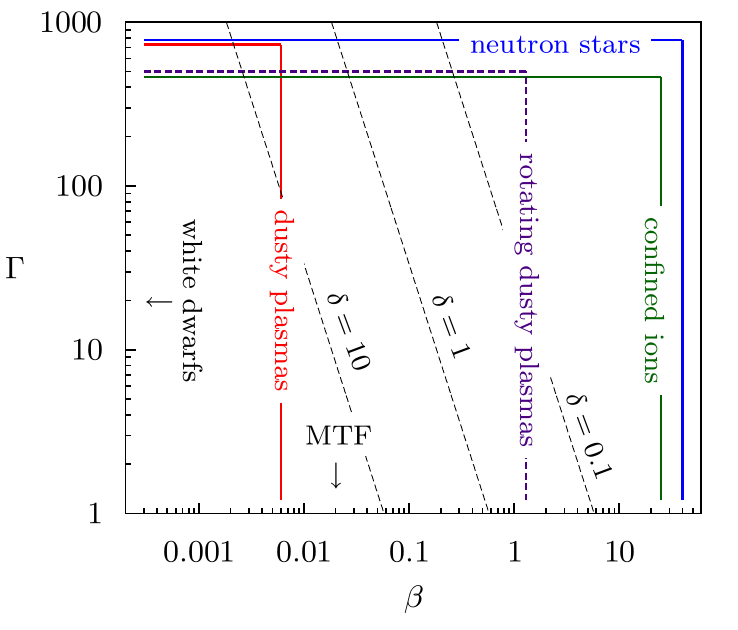}}
\caption{Approximate regimes of strongly correlated magnetized plasmas in the magnetization-coupling parameter plane. Lines of constant parameter $\delta$ [\eqref{eqn:delta} for $\text{dim}=3$] are shown together with important examples of such systems (MTF = magnetized target fusion~\cite{cereceda2008,lanl}).
}
\label{fig:n-t}
\end{figure}

If a magnetic field is turned on, the particles are, in addition to the Coulomb force, subject the Lorentz force giving rise to gyration around the field lines with the cyclotron frequency and the Larmor radius 
\begin{equation}\label{eqn:omega_c}
 \omega_\text{c} = \frac{QB}{m}, \qquad r_\text{L} = \frac{v_T}{\omega_\text{c}}.
\end{equation}
The effect on the particle trajectories is seen in figure~\ref{fig:traject}. In the middle column B-field and correlations have a comparable effect on the trajectories whereas, in the right column, the B-field clearly dominates, giving rise to the familiar helical orbits. At strong coupling, evidently, the orbits are much smaller, nevertheless the particles remain localized around their mean positions. We may quantify the relative strength of correlations and magnetic field by introducing dimensionless energy and length parameters,
\begin{eqnarray}
 \beta &=&\frac{\omega_\text{c}}{\omega_\text{p}} = B\sqrt{\frac{4\pi \epsilon_0}{\text{dim}} \frac{a_\text{ws}^3}{m}},\label{eqn:beta}\\
\delta&=&\frac{r_\text{L}}{a_\text{ws}} = \frac{1}{\beta}\frac{1}{\sqrt{{\rm dim}\cdot\Gamma}}.\label{eqn:delta}
\end{eqnarray}
Typical parameter regimes for various strongly coupled plasmas are shown in figure~\ref{fig:n-t}.

\begin{figure}
\includegraphics[height=6cm]{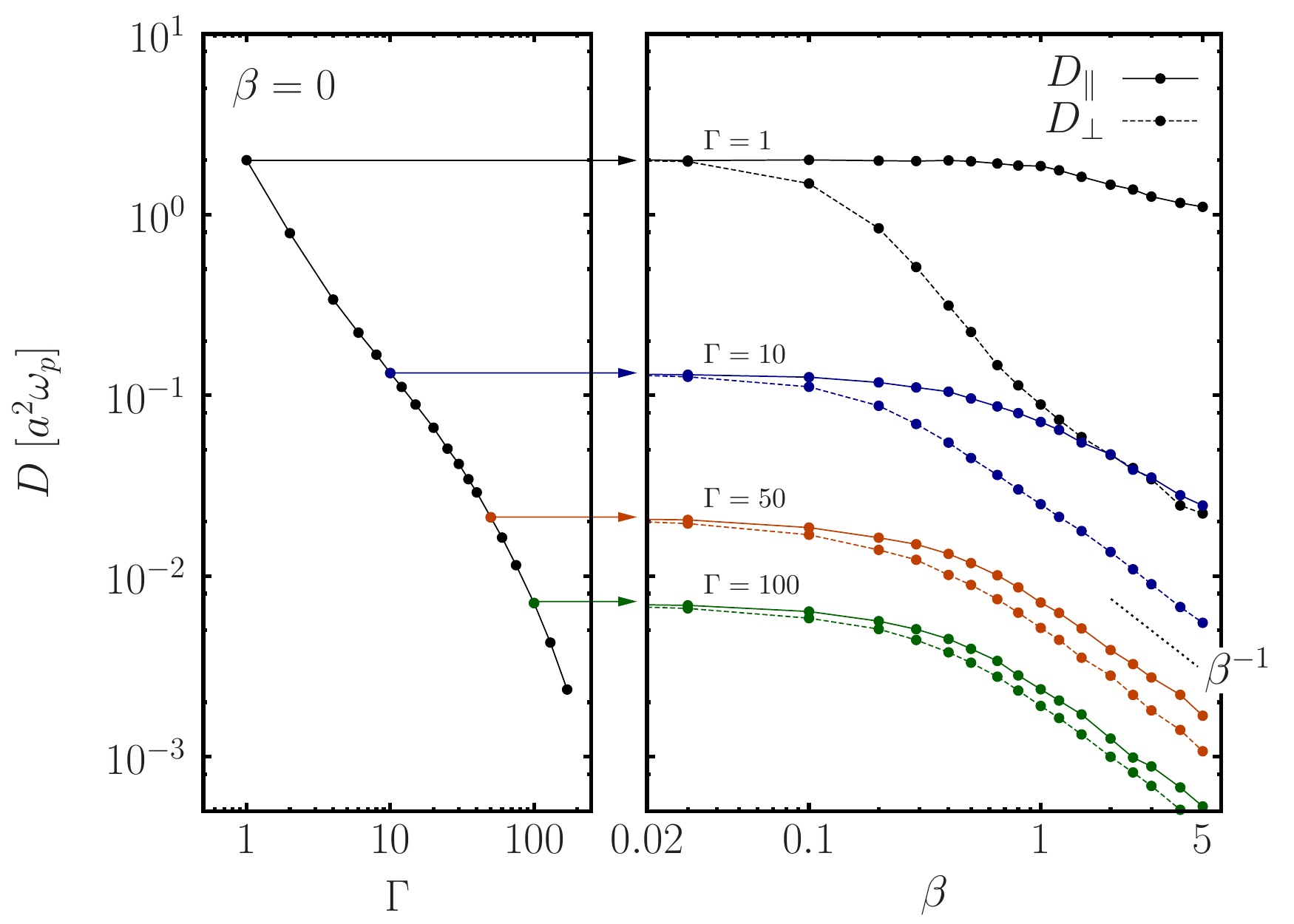}
\caption{Decay of the diffusion coefficient of a one-component plasma when coupling is increased (left) and when, additionally, the magnetization is increased (right). With data from~\cite{ott2011}.}\label{fig:d}
\end{figure}

From the trajectories it should be expected that macroscopic transport properties of the plasma will be strongly influenced both by Coulomb correlations and the magnetic field. An increase of $\Gamma$ leads to localization and thus tends to reduce the particle mobility. Similarly, an increase of the field strength will also tend to confine the particles in the direction perpendicular to ${\bf}B$. 
The combined effect of Coulomb correlations and magnetic field on the particle transport in a 3D OCP has recently been studied by first-principle MD simulations \cite{ott2011} and confirmed these expectations as demonstrated in figure~\ref{fig:d}. There is a systematic reduction of the diffusion coefficient with $\Gamma$, left figure. With increased magnetic field the diffusion perpendicular to ${\bf B}$ is further inhibited (right figure). An interesting effect is seen in the longitudinal diffusion coefficient. Although there is no component of the Lorentz force parallel to ${\bf B}$ and particle transport in this direction should remain unaffected, this picture breaks down in a strongly correlated plasma. The reason is that, upon increase of $\Gamma$, large-angle scattering occurs more frequently which couples the motion perpendicular and parallel to the field. As a result the diffusion parallel to ${\bf B}$ becomes also influenced (and suppressed) by the field, and eventually (see curves for $\Gamma=100$) the two diffusion coefficients approach each other. 

We also point out that the decay of both diffusion coefficients approaches, for large B-fields, a $1/B$ power law which is familiar from ideal high-temperature plasmas (Bohm diffusion). This indicates that here collective mechanisms such as transport mediated by magneto-plasmon excitations become dominant. In fact, the oscillation spectrum of magnetized strongly correlated plasmas has recently been computed numerically for 2D \cite{bonitz_prl10, ott_pre11} and 3D systems \cite{ott_prl12}. The magnetic field was found to drastically alter the oscillation spectrum giving rise to nonlinear effects such as Bernstein modes and higher harmonics. Again, these effects become relevant for sufficiently strong fields characterized by $\beta \gtrsim 0.5$.

Summarizing, we note that correlations and magnetic field (for $\beta\le 5$) suppress the particle mobility by about four orders of magnitude compared to the case of an unmagnetized weakly coupled plasma which is crucial for a broad range of laboratory plasma applications and for many astrophysical problems, such as the evolution of stellar objects. It would, therefore, be highly desirable to probe this regime of strongly correlated and strongly magnetized plasmas with $\Gamma \gg 1$ and $\beta \gtrsim 0.5$ in a laboratory experiment with dusty plasmas making use of the unprecedented resolution capabilities, e.g. \cite{ivlev2012,bonitz2010}. After extreme values of $\Gamma$ are now routinely achieved, let us now estimate the possible values of $\beta$. Assuming a mass of $m\approx 10^{-13}\,\text{kg}$, one obtains for a typical experiment ($a_\text{ws}\approx 2.5\cdot 10^{-4}\,\text{m}$~\cite{liu2008}) and a field of $B=4$T, common for superconducting magnets, a magnetization $\beta\approx 4\cdot 10^{-4}$. This value may be increased if one uses smaller (i.e., lighter) particles. However, even then, this is still below the range of interest. Furthermore, one would lose the key advantages of dusty plasmas: first, smaller particles collect a lower charge which decreases $\Gamma$. Second, the diagnostics of very small particles becomes increasingly difficult, and more sophisticated methods other than video microscopy may become necessary to reliably detect the dust grains. Finally, while the dust is practically un-magnetized, electrons and (partially) ions are magnetized resulting in very different plasma properties.

\section{Rotating dusty plasmas}\label{s:rotation}
A way out of this seemingly hopeless situation has been recently proposed and demonstrated in~\cite{kaehlert2012}. The key is the fourth component of a complex plasma---the neutral gas---which has been completely disregarded so far in our discussion. Even though the typical dust-neutral damping rates are very low, in particular compared to colloidal suspensions~\cite{ivlev2012}, the small yet finite friction force is sufficient to control the global motion of the dust component. This was demonstrated in experiments by Carstensen \textit{et al.}~\cite{carstensen2009}, who used a spinning electrode to create a global rotation of the neutral gas in the discharge chamber. One application of their method involved the centrifugal force, which was used to determine the particle charge~\cite{carstensen2010}. It was recently demonstrated~\cite{kaehlert2012} that the second apparent force in a rotating reference frame---the Coriolis force---can be used to ``magnetize'' the dust grains. We now present details of this approach.
\subsection{Effective magnetization}
In the laboratory system, the Langevin equation for the dust grains---being at positions $\bi r_i=(x_i,y_i,z_i)$, interacting via the force $\bi F_\text{int}({\bi r_{ij}})=-\nabla_{i} \phi(\rho_{ij},z_{ij})$ and confined in a potential $V(\rho,z)$---reads~\cite{kaehlert2012}
\begin{equation}\label{eqn:langevinEOM}
m\ddot{\bi r}_i(t) =-\nabla_i V(\rho_i,z_i) + \sum_{j\ne i}^N \bi F_\text{int}({\bi r_{ij}})-\nu m\left[ \dot{\bi r}_i(t) - \bi u({\bi r}_i) \right] + \bi f_i(t).
\end{equation}
Here, $\phi(\rho,z)$ is the interaction potential, $\rho=\sqrt{x^2+y^2}$, $z_{ij}=z_i-z_j$, $\rho_{ij}=\sqrt{x_{ij}^2+y_{ij}^2}$, $\bi u(\bi r)$ the flow field of the neutral gas, and $\nu$ the damping rate. The random force $\bi f_i(t)$ is characterized by $\langle \bi f_i(t)\rangle=0$ and $\langle f_i^\alpha(t) f_j^\beta(t') \rangle = 2m\nu k_\text{B} T_\text{n} \delta_{ij} \delta^{\alpha\beta} \delta(t-t')$, where $\alpha,\beta\in\{x,y,z\}$ [neutral gas temperature $T_\text{n}$]. For the special case of a uniform angular velocity $\Omega$, $\bi u(\bi r)=(\Omega\, \bi e_z)\times\bi r$, one can adopt the point of view of an observer at the origin of the coordinate system who co-rotates with the gas column. The coordinates in the laboratory $\{\bi r(t)\}$ and the rotating frame $\{\bar{\bi r}(t)\}$ are connected via (e.g.,~\cite{hestenes1999})
\begin{eqnarray}
\bi r(t)&=&\bi R(t)\, \bar {\bi r}(t),\nonumber\\
\dot{\bi r}(t)&=& \bi R(t) \left[ (\Omega\,\hat{\bi e}_z)\times  \bar {\bi r}(t)  + \dot{\bar {\bi r}}(t) \right],\label{eqn:trafo} \\
\ddot{\bi r}(t)&=& \bi R(t)\left[ (\Omega\,\hat{\bi e}_z)\times \left\{ (\Omega\,\hat{\bi e}_z)\times  \bar {\bi r}(t) \right\}   \right.\left.  + (2\,\Omega\,\hat{\bi e}_z)\times \dot{\bar {\bi r}}(t) + \ddot{\bar {\bi r}}(t) \right]\nonumber,
\end{eqnarray}
where the rotation matrix reads
\begin{equation}
   \bi R(t)=\begin{pmatrix}
               \cos(\Omega t) & -\sin(\Omega t) & 0\\ 
   	    \sin(\Omega t) & \cos(\Omega t) & 0\\
               0 & 0 & 1
             \end{pmatrix}. 
\end{equation}
Using~\eqref{eqn:trafo} in~\eqref{eqn:langevinEOM} and multiplying by $\bi R^T(t)$ yields~\cite{kaehlert2012}
\begin{equation}\label{eqn:rot}
 m\ddot{\bar{\bi r}}_i(t) = -\bar\nabla_i \bar V(\bar\rho_i,\bar z_i) + \sum_{j\ne i}^N \bar{\bi F}_\text{int}(\bar{\bi r}_{ij}) + \bar{\bi F}_\text{Cor}(\dot{\bar{{\bi r}}}_i)-\nu m \dot{\bar{\bi r}}_i(t) + \bar{\bi f}_i(t),
\end{equation}
where $\bar{\bi F}_\text{Cor}(\dot{\bar{{\bi r}}})=2m\, \dot{\bar{{\bi r}}}\times (\Omega\, \bi e_z)$ denotes the Coriolis force.

Since the angular rotation frequency of the gas is constant, eq.~\eqref{eqn:rot} has the same form as the Langevin equation for a stationary gas. The confinement potential in the rotating frame must be complemented by the centrifugal potential, i.e., 
\begin{align}
\bar V(\bar\rho,\bar z)=V(\bar\rho,\bar z)-\frac{m}{2}\Omega^2 \bar\rho^2
 \end{align}
whereas the interaction forces are equivalent to those in the laboratory frame. The random force is given by $\bar{\bi f}_i(t)= \bi R^T(t)\bi f_i(t)$. While $\bar f_i^z(t)=f_i^z(t)$, the $x$ and $y$ components require more attention. As an example, we consider $\bar f_i^x(t)=\cos(\Omega t) f_i^x(t)+\sin(\Omega t) f_i^y(t)$ and calculate its properties. Since $\langle f_i^x(t)\rangle=\langle f_i^y(t)\rangle=0$ and both $f_i^x(t)$ and $\,f_i^y(t)$ are not correlated with $f_i^z(t)$, we have $\langle\bar f_i^x(t)\rangle=0$ and $\langle \bar f_i^x(t) \bar f_i^z(t') \rangle=0$, respectively. Using $\langle f_i^x(t)f_i^y(t') \rangle=0$ and $\langle f_i^x(t) f_i^ x(t') \rangle=\langle f_i^y(t) f_i^y(t') \rangle=2m\nu k_\text{B} T_\text{n} \delta(t-t')$, one can further show that $\langle \bar f_i^x(t) \bar f_i^y(t') \rangle=0$ and $\langle \bar f_i^x(t) \bar f_i^x(t') \rangle=2m\nu k_\text{B} T_\text{n} \delta(t-t')$. The calculations for the $y$ component are analogous, and the results can be summarized as $\langle\bar{\bi f}_i(t)\rangle=0$ and $\langle \bar{f}_i^\alpha(t) \bar f_j^\beta(t') \rangle = 2m\nu k_\text{B} T_\text{n} \delta_{ij} \delta^{\alpha\beta} \delta(t-t')$, i.e., the mean and the correlation function of the random force in the laboratory and the rotating frame are equivalent.

The mathematical equivalence between the Lorentz force $\bi{F}_{\text{\scriptsize{L}}}(\dot{{{\bi r}}})=Q\,\dot{{{\bi r}}}\times (B\, \bi e_z)$ and the Coriolis force is the reason why the approach can be used to study ``magnetized'' plasmas. When observed in the rotating frame and taking the centrifugal potential into account, the dust particles follow the same physical laws as in a constant magnetic field $\bi B_\text{eff}=(2m  \Omega/Q)\hat{\bi e}_z$. The important quantity for the magnetization parameter~\eqref{eqn:beta} is the cyclotron frequency $\omega_\text{\scriptsize{c}}$, which can now be replaced by $2\Omega$. Since rotation frequencies of a few Hz are easily reached, strong magnetization of the dust component becomes possible, as was verified in experiments with a small cluster~\cite{kaehlert2012}. 

Thus rather than trying to overcome the problems encountered when using expensive superconducting magnets, we exploit the advantages a dusty plasma offers. The coordinate transformation from the laboratory to the rotating frame is easily performed, and the same data analysis techniques as for a stationary plasma can be applied. The variation of the plasma parameters (coupling, screening) induced by the centrifugal force are negligible for small rotation frequencies and can be directly accounted for at higher rotation speeds. While the scaling relations for small systems have already been discussed~\cite{kaehlert2012}, we now focus on larger ensembles and the transition to bulk behaviour.

\subsection{Parameter variation in bulk systems}
Particle confinement in dusty plasma experiments is provided by various forces including gravity, electric fields and thermophoresis. In the following, we consider a two-dimensional system (implying strong vertical confinement) and assume that the particles interact via the Yukawa potential $\phi(r)=(Q^2/4\pi\epsilon_0\,r) \exp(-r/\lambda)$, where $\lambda$ denotes the Debye length. The effective in-plane confinement is given by a harmonic potential $V(r)=m\omega_\perp^2 r^2/2$, and all higher order terms are neglected. Their contribution may become important for large cluster sizes or very high rotation frequencies. 

\begin{figure}

\includegraphics{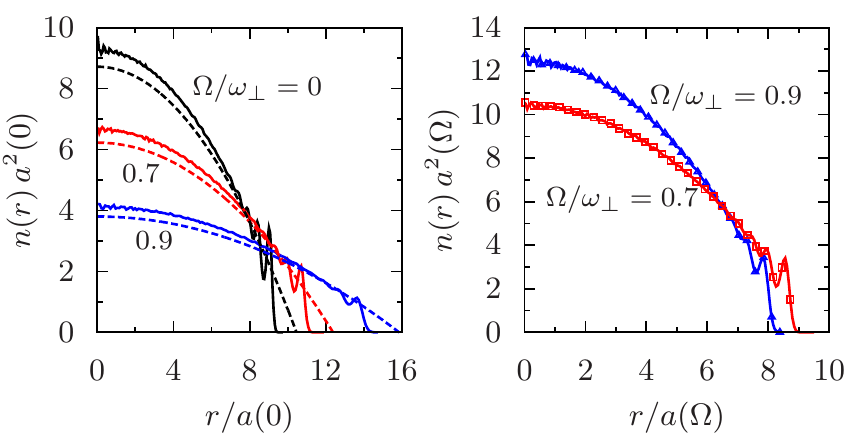}
\caption{Radial density profile of $N=1500$ particles in a rotating gas flow. {\bf  Left}: Radial density for $\Omega/\omega_\perp=0,\, 0.7,\, 0.9$ with $\kappa(0)=1$ and $\Gamma(0)=10$. The dashed lines show a theoretical result for $\Gamma\to\infty$ derived by Totsuji \textit{et al.}~\cite{totsuji2001} [equation~(5.17) in their paper], where the screening parameter $\kappa(\Omega)$ been adjusted according to~\eqref{eqn:finitescale}. The length scale has been modified accordingly. {\bf Right}: Comparison of the density profile at finite rotation frequencies (lines) with results at $\Omega=0$ (symbols) and adjusted screening and coupling parameters. The wiggles at large radii are due to the onset of shell formation which is not included in the analytical model~\cite{totsuji2001}.}\label{fig:density}
\end{figure}

We first define the dimensionless screening parameter $\kappa(\Omega)=a(\Omega)/\lambda$, the coupling parameter $\Gamma(\Omega)=Q^2/(4\pi\epsilon_0 \,a(\Omega)\,k_\text{B} T_\text{n})$, and the friction coefficient $\gamma(\Omega)=\nu/\bar\omega(\Omega)$, where $a(\Omega)=[Q^2/4\pi \epsilon_0  m\bar\omega^2(\Omega)]^{1/3}$ and $\bar\omega(\Omega)=\sqrt{\omega_\perp^2-\Omega^2}$. They are useful dimensionless parameters for a finite system and scale as~\cite{kaehlert2012}
\begin{equation}\label{eqn:finitescale}
\frac{\kappa(\Omega)}{\kappa(0)}=\left(1-\frac{\Omega^2}{\omega_\perp^2} \right)^{-1/3}, \frac{\Gamma(\Omega)}{\Gamma(0)}=\left(1-\frac{\Omega^2}{\omega_\perp^2} \right)^{1/3}, \frac{\gamma(\Omega)}{\gamma(0)}=\left(1-\frac{\Omega^2}{\omega_\perp^2} \right)^{-1/2}.
\end{equation}
The scaling behaviour is illustrated in figure~\ref{fig:density}, where we show the radial density profile in a stationary gas and at finite rotation frequencies obtained from the numerical solution of the Langevin equations~\eqref{eqn:langevinEOM} with $N=1500$ dust particles~\footnote{We use the ``SLO'' integrator derived by Mannella in~\cite{Langevin}}. As shown in the left panel, the density drops as we increase the rotation frequency. If the screening and coupling parameters are adjusted according to~\eqref{eqn:finitescale}, we find excellent agreement with simulations at $\Omega=0$ (right panel), as expected. The Coriolis force does not affect the static properties of the system.

Experiments typically use large yet finite dust monolayers containing a few thousand particles to investigate bulk properties of dusty plasmas~\cite{nunomura2006, liu2008, hartmann2011}. The applicability of this approach was investigated theoretically by Sheridan~\cite{sheridan2007}. To give estimates for the scaling of the plasma parameters in bulk systems, we refer to the theoretical results obtained by Totsuji \textit{et al.}~\cite{totsuji2001}, who developed a theory for the average density profile of large clusters. One of their results, using the local density approximation and neglecting correlations, is shown in figure~\ref{fig:density} and shows good agreement with the simulation data. While the theory was derived for $\Gamma\to \infty$, it still provides a good description of the mean density for intermediate coupling. In agreement with the results of~\cite{totsuji2001} for crystallized plasmas, it slightly underestimates the density in the inner region of the trap. Larger deviations were observed for $\kappa^3\ll 1$ or $\kappa^3\gg 1$. In the latter case, the theory could be improved by taking the correlation energy into account~\cite{totsuji2001}.

\begin{figure}
\includegraphics{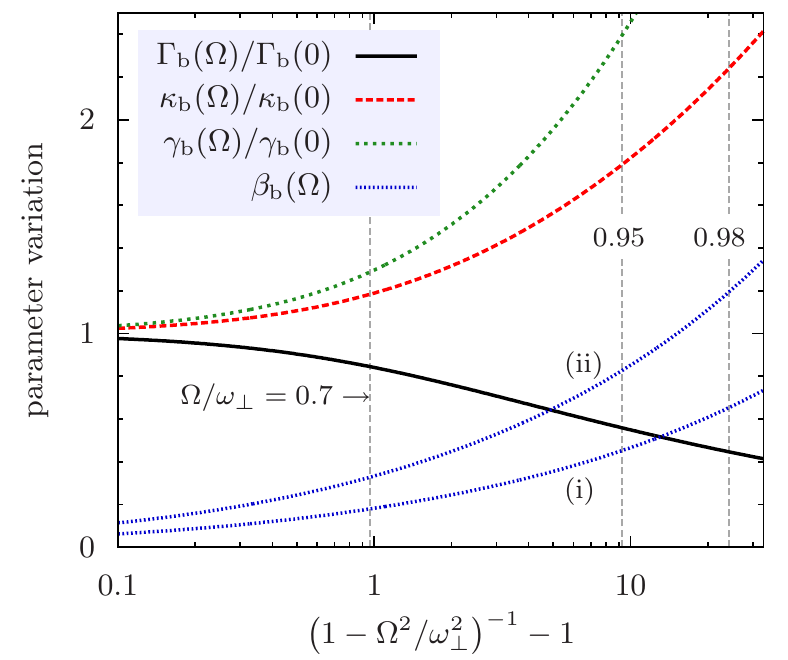}
\caption{Scaling of the coupling parameter, screening parameter, and damping constant in a 2D plasma according to~(\ref{eqn:bulkscale}). The magnetization $\beta_\text{b}$ is shown for (i) $N\kappa(0)=1500$ and (ii) $ N\kappa(0)=300$, }\label{fig:scaling}
\end{figure}

Based on their theory, Totsuji \textit{et al.}~\cite{totsuji2001} further give an estimate for the mean inter-particle spacing, $a_\text{ws}=\lambda(8\kappa^3/N)^{1/4}$, where the $\kappa^{3/4}$ scaling was found to be in good agreement with simulation results~\cite{sheridan2007}. This enables us to estimate the relevant dimensionless parameters for a bulk system ('b') rotating with angular velocity $\Omega$ as
\begin{eqnarray}\label{eqn:bulkscale}
\Gamma_\text{b}(\Omega)&=& \frac{Q^2/4\pi\epsilon_0}{a_\text{ws}(\Omega)\, k_\text{B} T_\text{n}}\approx \Gamma(0) \left[\frac{N \kappa(0)}{8}\right]^{1/4} \left(1-\frac{\Omega^2}{\omega_\perp^2} \right)^{1/4} ,\\[1ex]
\kappa_\text{b}(\Omega)&=&\frac{a_\text{ws}(\Omega)}{\lambda}\approx \left[\frac{8\kappa^3(0)}{N} \right]^{1/4}\left(1-\frac{\Omega^2}{\omega_\perp^2} \right)^{-1/4} ,\nonumber\\[1ex]
\gamma_\text{b}(\Omega)&=&\frac{\nu}{\omega_\text{p}(\Omega)}\approx \frac{\gamma(0)}{\sqrt{2}} \left[\frac{8}{N\kappa(0)} \right]^{3/8} \left(1-\frac{\Omega^2}{\omega_\perp^2} \right)^{-3/8}  ,\nonumber\\[1ex]
\beta_\text{b}(\Omega)&=&\frac{2\Omega}{\omega_\text{p}(\Omega)}\approx \sqrt{2}\left[\frac{8}{N\kappa(0)} \right]^{3/8} \frac{\Omega}{\omega_\perp} \left(1-\frac{\Omega^2}{\omega_\perp^2} \right)^{-3/8}.\nonumber
\end{eqnarray}
The exponents are very similar to those in~\eqref{eqn:finitescale}, but the variation of the screening and coupling parameters with $\Omega$ is slightly reduced. The trend of eqs.~(\ref{eqn:bulkscale}) is displayed in figure~\ref{fig:scaling}.
Considering the magnetization $\beta_\text{b}(\Omega)$, the factor $N^{-3/8}$ indicates that it could be more difficult to achieve strong magnetization in large extended systems than in small clusters. For the parameters in figure~\ref{fig:density}, we obtain $\Gamma_\text{b}(0)\approx 37$ and $\kappa_\text{b}(0)\approx 0.27$, which drop to $\Gamma_\text{b}\approx 21$ and increase to $\kappa_\text{b}\approx 0.48$ at $\Omega/\omega_\perp=0.95$. The magnetization at this rotation frequency is $\beta_\text{b}\approx 0.45$, which is sufficient to observe the magnetic field effects discussed in Sec.~\ref{s:magnetization}. In particular, this should allow to clearly detect the magnetoplasmon and magnetoshear modes~\cite{jiang2007,hou2009}, and could even be sufficient to observe higher harmonics of the former~\cite{bonitz_prl10, ott_pre11}. The actual scaling behaviour in experiments could differ from these predictions due to anharmonic contributions to the confinement potential or deviations of the interaction from a Yukawa potential. Further, the limitations of the theory~\cite{totsuji2001} must be kept in mind.

\section{Conclusions}
In this paper we have given an overview on the combined effect of strong Coulomb correlations and strong magnetization. The main trend is a reduction of particle mobility which should play an important role for the transport properties of such plasmas. Relevant examples have been presented in figure~\ref{fig:n-t}.  Dusty plasmas provide a unique opportunity to accurately study correlation effects, due the large particle size. At the same time, this prevents a magnetization of the dust particles, even in the strongest fields superconducting magnets can deliver. Instead, strong ``magnetization'' of the heavy dust particles can be achieved by putting them into rotation by using a rotating neutral gas column. The appearing Coriolis force is equivalent to the Lorentz force, and the system mimics one in a strong magnetic field that is of the order of $10^4~\text{T}$. This concept was recently demonstrated experimentally for a dust cluster of $N=4$ particles~\cite{kaehlert2012}. Here, we have performed first-principle Langevin dynamics simulations that confirm the scaling of the plasma parameters in a finite system. Based on theoretical predictions for the density profile~\cite{totsuji2001}, we have derived the scaling of the relevant plasma parameters with the rotation frequency in a bulk system, which should allow for a direct comparison with experiments.

\ack
This work was supported by the DFG via SFB TR24 and SFB TR6, the ERC Advanced Grant INTERCOCOS and a grant for CPU time at the HLRN.

\section*{References}

\bibliographystyle{unsrt}

\begin{thebibliography}{10}

\bibitem{ivlev2012}
A.~V. Ivlev, H.~L{\"o}wen, G.~E. Morfill, and C.~P. Royall.
\newblock {\em Complex plasmas and colloidal dispersions: particle-resolved
  studies of classical liquids and solids}.
\newblock Series in Soft Condensed Matter - Vol. 5. World Scientific, 2012.

\bibitem{bonitz2010}
M.~{Bonitz}, C~Henning, and D.~Block.
\newblock {Complex plasmas: a laboratory for strong correlations}.
\newblock {\em Rep. Prog. Phys.}, 73:066501, 2010.

\bibitem{potekhin2010}
Aleksandr~Y Potekhin.
\newblock The physics of neutron stars.
\newblock {\em Physics-Uspekhi}, 53(12):1235, 2010.

\bibitem{ott2011}
T.~Ott and M.~{Bonitz}.
\newblock {Diffusion in a strongly coupled magnetized plasma}.
\newblock {\em Phys. Rev. Lett.}, 107:135003, 2011.

\bibitem{bonitz_prl10}
M.~Bonitz, Z.~Donk\'o, T.~Ott, H.~K\"ahlert, and P.~Hartmann.
\newblock {Nonlinear Magnetoplasmons in Strongly Coupled Yukawa Plasmas}.
\newblock {\em Phys. Rev. Lett.}, 105:055002, 2010.

\bibitem{ott_prl12}
T.~Ott, H.~K\"ahlert, A.~Reynolds, and M.~Bonitz.
\newblock {Oscillation Spectrum of a Magnetized Strongly Coupled
  One-Component-Plasma}.
\newblock {\em Phys. Rev. Lett.}, 108:255002, 2012.

\bibitem{kaehlert2012}
H.~{K{\"a}hlert}, J.~{Carstensen}, M.~{Bonitz}, H.~{L{\"o}wen}, F.~{Greiner},
  and A.~{Piel}.
\newblock {Magnetizing a complex plasma without a magnetic field}.
\newblock {\em Phys. Rev. Lett.}, in press.
\newblock {\em ArXiv e-prints}, 1206.5073 2012.

\bibitem{cereceda2008}
C.~Cereceda, M.~DePeretti, and C.~Deutsch.
\newblock Thermal conductivity coefficient and stopping power in magnetized
  target fusion (mtf) scenario.
\newblock In E.~Panarella and R.~Raman, editors, {\em Current Trends in
  International Fusion Research: Proceedings of the Fifth Symposium}, page 155.
  NRC Research Press, 2008.

\bibitem{lanl}
http://wsx.lanl.gov/mtf.html/.
\newblock Magnetized target fusion experiments at lanl.
\newblock 2012.

\bibitem{ott_pre11}
T.~Ott, M.~Bonitz, P.~Hartmann, and Z.~Donk\'o.
\newblock Higher harmonics of the magnetoplasmon in strongly coupled coulomb
  and yukawa systems.
\newblock {\em Phys. Rev. E}, 83:046403, 2011.

\bibitem{liu2008}
Bin Liu and J.~Goree.
\newblock Superdiffusion and non-gaussian statistics in a driven-dissipative 2d
  dusty plasma.
\newblock {\em Phys. Rev. Lett.}, 100:055003, 2008.

\bibitem{carstensen2009}
J.~{Carstensen}, F.~{Greiner}, L.-J. {Hou}, H.~{Maurer}, and A.~{Piel}.
\newblock {Effect of neutral gas motion on the rotation of dust clusters in an
  axial magnetic field}.
\newblock {\em Phys. Plasmas}, 16(1):013702, 2009.

\bibitem{carstensen2010}
J.~{Carstensen}, F.~{Greiner}, and A.~{Piel}.
\newblock {Determination of dust grain charge and screening lengths in the
  plasma sheath by means of a controlled cluster rotation}.
\newblock {\em Phys. Plasmas}, 17(8):083703, 2010.

\bibitem{hestenes1999}
D.~Hestenes.
\newblock {\em New foundations for classical mechanics}.
\newblock Fundamental theories of physics. Kluwer Academic Publishers, 1999.

\bibitem{Langevin}
R.~Mannella.
\newblock Quasisymplectic integrators for stochastic differential equations.
\newblock {\em Phys. Rev. E}, 69(4):041107, 2004.

\bibitem{totsuji2001}
Hiroo Totsuji, Chieko Totsuji, and Kenji Tsuruta.
\newblock Structure of finite two-dimensional yukawa lattices: Dust crystals.
\newblock {\em Phys. Rev. E}, 64:066402, 2001.

\bibitem{nunomura2006}
S.~Nunomura, D.~Samsonov, S.~Zhdanov, and G.~Morfill.
\newblock Self-diffusion in a liquid complex plasma.
\newblock {\em Phys. Rev. Lett.}, 96:015003, 2006.

\bibitem{hartmann2011}
Peter Hartmann, M\'at\'e~Csaba S\'andor, Anik\'o Kov\'acs, and Zolt\'an
  Donk\'o.
\newblock Static and dynamic shear viscosity of a single-layer complex plasma.
\newblock {\em Phys. Rev. E}, 84:016404, 2011.

\bibitem{sheridan2007}
T.~E. {Sheridan}.
\newblock {Criterion for bulk behavior of a Yukawa disk}.
\newblock {\em Physics of Plasmas}, 14(3):032108, 2007.

\bibitem{jiang2007}
Ke~Jiang, Yuan-Hong Song, and You-Nian Wang.
\newblock Theoretical study of the wave dispersion relation for a
  two-dimensional strongly coupled yukawa system in a magnetic field.
\newblock {\em Phys. Plasmas}, 14(10):103708, 2007.

\bibitem{hou2009}
Lu-Jing Hou, P.~K. Shukla, Alexander Piel, and Z.~L. Mi\v{s}kovi\'{c}.
\newblock Wave spectra of two-dimensional yukawa solids and liquids in the
  presence of a magnetic field.
\newblock {\em Phys. Plasmas}, 16(7):073704, 2009.

\end{thebibliography}

\end{document}